\documentstyle[preprint,eqsecnum,aps]{revtex}
\begin{document}
\draft
\preprint{HYUPT-94/04 SNUTP 94-64\hspace{-26.5mm}
\raisebox{2.4ex}{hep-th/9409006}}
\title{Quantum Aspects of Supersymmetric Maxwell Chern-Simons Solitons}
\author{Bum-Hoon Lee}
\address{
Department of Physics, Hanyang University, Seoul 133-791, Korea}
\author{Hyunsoo Min}
\address{
Department of Physics, Seoul City University, Seoul 130-743, Korea}
\date{\today}
\maketitle
\begin{abstract}
We study the various quantum aspects of the $N=2$ supersymmetric Maxwell
Chern-Simons vortex systems. The fermion zero modes around
the vortices will give rise the degenerate states of vortices.
We analyze the angular momentum of these zero modes and apply the result
to get the supermultiplet structures of the vortex.
The leading quantum correction to the mass of the vortex coming from the
mode fluctuations is also calculated using various methods depending on
the value of the coefficient of the Chern-Simons term $\kappa$ to be zero,
infinite and finite,
separately.
The mass correction is
shown to vanish for all cases. Fermion numbers of vortices
are also discussed.
\end{abstract}

\section{Introduction}\label{sec:1}
Abelian gauge theory in 2+1 dimensions with the Chern-Simons (CS)
term\cite{C-Sterm,C-Sspin} has
attracted much interest.  Matter fields coupled with this term is believed
to describe the anyons with the fractional spin and fractional statistics.
Such important planar phenomena as the high $T_c$ superconductivity and
the fractional quantum Hall effect has added more interest in the field
theory models with CS term. The characters of allowed solitons
are also affected by the presence of the CS term.  As is well
known, the usual 2+1 dimensional Abelian Higgs model supports only
electrically neutral vortices as topologically stable soliton
solutions\cite{Vortex}. On the other hand, the CS term makes
the vortices\cite{CSvortex} electrically charged, which
are (extended) anyons\cite{anyon}. We have quite rich vortex structures
depending on whether the matter fields are relativistic \cite{CSHiggs} or
non-relativistic\cite{JP} and whether we have more than one CS
fields\cite{multiCS}.
In this work, we are mainly interested in the case with the relativistic
matter fields coupled with the gauge field with both Maxwell and
CS term in general. As a special limit of this general model, we
get the Abelian Higgs model and the `minimal' Chern-Simons Higgs model
(i.e., without the Maxwell term in the action).

With some special choice of the scalar potential in (2+1)-dimensional
gauge models Ref.\cite{MCS}, one can obtain interesting limiting theories
in which the
minimum energy static soliton solutions satisfy first-order differential
equations, called the Bogomol'nyi\cite{Bogo} or self-duality equations.
This special potential becomes a specific scalar quartic
potential for the Abelian Higgs model, while in the case of the `minimal'
Chern-Simons Higgs model a specific sixth-order potential form
\cite{CSHiggs,JLW90}.  The appearance
of self-dual structures for certain special Higgs potentials can be
ascribed to the extended supersymmetry\cite{diVecchiaF,SUSY,Ferzeromode}.
Requiring
an $N=2$ supersymmetry guarantees this special form of the potential.
There also exists an $N=1$ supersymmetric model which produces exactly the
same bosonic part of the Lagrangian as that of the $N=2$ model. The
fermion number is, however, not preserved in this case. We will mainly
consider the model with more symmetry, i.e., the model with the $N=2$
supersymmetry.

A remarkable feature with these self-dual systems is the existence of
static multi-vortex solutions which represent static configurations of
vortices with unit flux without any interaction energy between them. This
interpretation is supported by counting independent zero
modes\cite{JLW90,LeeMinRim}to the
boson fluctuation equations in the background field of a particular
soliton solution. These bosonic zero modes are related with the collective
modes of the solitons and play an important role in
understanding the dynamics of the slowly moving vortices \cite{slowmotion}.

Fermions around the vortices also have zero modes.
For the models under study it is found that {\em all} fermion zero modes
around the general multi-vortex background are closely related to the
corresponding bosonic zero modes.  The $N=2$ supersymmetry is crucial
\cite{Ferzeromode}.
They are very important
in the quantum study of the models, representing the degeneracy of the
soliton states (in contradistinction to bosonic zero modes which become
collective coordinates)\cite{JackiwRebbi}.  In supersymmetric models in
particular, they account for the soliton supermultiplet
structure\cite{susymonopole}.

The organization of this paper is as follows.
In Sec. II, we review the properties of the  $N=2$ supersymmetric
Maxwell Chern-Simons system; The particle contents, the properties of the
vortices, the symmetries and the corresponding algebra of the Lagrangian
are reviewed.
In Sec.III, we will quantize the fluctuating modes around the self-dual
vortices. By analysing the angular momentum fermion zero modes, we
describe the multiplet structures of the vortices.  Spin contents of the
degenerate supermultiplet of the vortex states are also calculated. Then,
in Sec. IV, we calculate the mass correction to the
vortices. We will do
this for the value of $\kappa$ to be zero, infinite and finite,
separately. Section V contains the summary of
our work and discussions.  Some technical details related with the
spin assignment, supermultiplets and phase shift analysis are described in
the appendices.
\section{Supersymmetric Maxwell Chern-Simons theory}\label{sec:2}

The Lagrangian for the Maxwell-Chern-Simons system with $N=2$
supersymmetry is given by \cite{Ferzeromode}
\begin{equation}
{\cal L} = {\cal L}_{B} + {\cal L}_{F},  \label{Lagtot}
\end{equation}
where
\begin{eqnarray}
{\cal L}_{B}&=&-{1\over 4} F_{\mu\nu} F^{\mu\nu}
+ {\kappa\over 4} \epsilon^{\mu\nu\lambda} F_{\mu\nu} A_\lambda
-|D_\mu \phi|^2 -{1\over2}(\partial_\mu N)^2  \nonumber \\
& &-{1\over2}(e|\phi|^2 +\kappa N -e v^2)^2 -e^2 N^2 |\phi|^2,
\label{Boson}
\end{eqnarray}
and
\begin{eqnarray}
{\cal L}_{F} &=&  i\bar\psi \gamma^\mu D_\mu \psi
 + i\bar\chi \gamma^\mu \partial_\mu  \chi  +{\kappa}
   \bar\chi\chi \nonumber\\
&&-i\sqrt{2}e(\bar \psi \chi \phi -\bar\chi \psi \phi^*)
 +eN\bar\psi\psi . \label{Fermion2}
\end{eqnarray}
Here, $D_\mu = \partial_\mu -ieA_\mu $ is the covariant derivative,
$N$ a real scalar, $\phi$ a complex charged scalar, and
$\psi$ ($\chi$) is a complex charged (neutral) 2-component spinor.
Our metric tensor $\eta^{\mu\nu}$ has the signature $(-,+,+)$. We will
choose the
$\gamma$-matrices as $\gamma^\mu  = (\sigma_3,i\sigma_2,i\sigma_1)$.

When the coupling strength $\kappa$ for the Chern-Simons term becomes zero,
the above Lagrangian reduces
to the $N=2$ supersymmetric abelian Higgs model\cite{diVecchiaF}. The
scalar potential in this limit
allows only the symmetry broken vacuum.
In another extreme limit of very large $\kappa$ (with the ratio
${e^2\over\kappa}$ fixed), the neutral scalar field $N$ (spinor field
$\chi$) can be represented in terms of the complex scalar field $\phi$ (spinor
field $\psi$) as
\begin{equation}
N= -{1\over \kappa} e(|\phi|^2-v^2), \quad
\chi = -{i\over\kappa}\sqrt{2} e \phi^* \psi,  \label{solauxN2}
\end{equation}
and the Lagrangian becomes the supersymmetric extension of the minimal
self-dual Chern-Simons Higgs model given in Ref.\cite{SUSY},
\begin{eqnarray}
{\cal L}_{\rm CS}^{(2)} &=& {\kappa\over 4} \epsilon^{\mu\nu\lambda}
F_{\mu\nu} A_\lambda -|D_\mu \phi|^2 -{e^4\over \kappa^2}|\phi|^2
(|\phi|^2 - v^2)^2 \nonumber \\ &&+ i\bar\psi \gamma^\mu D_\mu \psi
-{e^2\over\kappa} (3|\phi|^2 -v^2) \bar\psi\psi \label{Linfty}.
\end{eqnarray}

The above theory posesses various kinds of symmetries and their
corresponding currents. The energy-momentum vectors related with the
translational symmetry is given by
\begin{equation}
P^\mu = \int d^2  x \Theta^{0\mu}
\end{equation}
with the energy momentum tensor given by
\begin{equation}
\Theta_{\mu\nu} = F_{\mu\lambda}F_{\nu}{}^\lambda +D_\mu\phi^*D_\nu\phi
+D_\nu\phi^*D_\mu\phi +\partial_\mu N \partial_\nu N
-i\bar\psi\gamma_{(\mu} D_{\nu)}\psi -i\bar\chi\gamma_{(\mu}
\partial_{\nu)} \chi +\eta_{\mu\nu}{\cal L} .
\end{equation}
This energy momentum tensor is gauge invariant corresponding to the
translation supplemented with an appropriate gauge transformation.
The generators for the Lorentz symmetry can be also found. For example,
the canonical angular momentum operator $J$ is given by
\begin{eqnarray}
 J &=& \int d^2 x [\epsilon_{ij}x^iP_j
-{1\over2}(\bar\psi\psi+\bar\chi\chi)] \nonumber \\
&=& J_B +J_F \label{angmomform}
\end{eqnarray}
with the contribution from the bosonic fields
\begin{equation}
J_B  =  \int d^2 x \epsilon_{ij} x^i [\partial^0 N \partial^j N
+D^0\phi^*D^j\phi +D^j\phi^*D^0\phi
+F^{0k}F^j{}_k],  \label{ang_bos}
\end{equation}
and that from fermions
\begin{equation}
J_F = -i\int d^2x  \epsilon_{ij} x^i [ \bar\psi\gamma^0 D_j\psi
+\bar\chi\gamma^0\partial_j\chi]
-{1\over2} \int d^2x (\bar\psi\psi +\bar\chi\chi) . \label{ang_fer}
\end{equation}

The theory in Eq.(\ref{Lagtot}) possesses the following supersymmetry:
\begin{eqnarray}
&&\delta_\eta A_\mu
=i (\bar\eta\gamma_\mu\chi-\bar\chi\gamma_\mu\eta) ,\nonumber\\
&&\delta_\eta\phi = \sqrt{2} \bar \eta \psi,\qquad \delta_\eta N
= i(\bar\chi\eta -\bar\eta \chi) ,\nonumber\\
&&\delta_\eta \psi = -\sqrt{2}
(i\gamma^\mu \eta D_\mu \phi - \eta F ) ,\nonumber\\
&&\delta_\eta \chi = \gamma^\mu \eta (\partial_\mu N
 -f_\mu) +
  i\eta G,       \label{SUSY2}  \label{susytransf}
\end{eqnarray}
where
\begin{equation}
F=eN \phi, \quad f_\mu=-{i\over2}\epsilon_{\mu\nu\lambda}F^{\nu\lambda},
\quad G=e|\phi|^2 +\kappa N -e v^2 . \label{FfmuG}
\end{equation}
Here the spinor parameter $\eta$ should be taken as being
{\em complex} Grassmannian.
The corresponding supercharges are
\begin{eqnarray}
Q &=&\sqrt{2}\int d^2x \bigl[
(D_\mu\phi)^*\gamma^\mu\gamma^0\psi
-\sqrt{2}iF^*\gamma^0 \psi  \nonumber \\
&&-i(\partial_\mu N +f_\mu)
\gamma^\mu\gamma^0\chi
-G\gamma^0\chi \bigr].           \label{supercharge}
\end{eqnarray}
The algebra of these with the supercharges are given by
\begin{equation}
\{Q_\alpha, \, \bar Q^\beta \} = 2\gamma_{\alpha}^{\mu\beta} P_\mu
-2 \delta^{\beta}_{\alpha} ev^2 \Phi, \label{susyalgebra}
\end{equation}
with $\Phi= \int d^2 x B$.

The particle spectrum will form representations of the above symmetry
algebra. General structure of supermultiplet and the spin assignment will be
described in the appendix.
In the case of $\kappa = 0$ where there is only symmetry broken vacuum,
all the particles are massive with the same mass
$\sqrt{2}|ev|$. We  have two (massive) vector modes with spin 1
and $-1$, two real
scalar modes of spin 0 and four spinor modes with two of spin 1/2 and the
other two of spin $-1/2$. For the fermions, the sign of the mass term will
depend on the spin. Two sets of the $N=2$ supermultiplets are formed. One
set is with one of spin 1, two of spin 1/2 and one of spin 0. The other is
with one of spin $-1$, two of spin $-1/2$ and one of spin 0.

For the Chern-Simons theory, the potential allows both symmetry unbroken
and symmetry broken vacua.  In
the symmetry broken vacuum, we have four degrees of freedom with the equal
masses $2e^2v^2/|\kappa|$ forming one set of $N=2$ supermultiplet. The
spin contents for $\kappa > 0$ are one of spin $-1$, two of spin $-1/2$ and
one of spin 0.  With $\kappa < 0$, all the spins in the supermultiplet
will change signs.  In the unbroken vacuum sector, all the four modes are
massive again with masses equal to $e^2v^2/|\kappa|$. These are split into
two supermultiplets. One supermultiplet consists with each of spin 0 and
spin 1/2, and the other with each of spin 1/2 and 1. If the sign of
$\kappa$ becomes negative, the signs of all the spins are also changed.

In the general case with finite value of $\kappa$, there are also two
degenerate ground states i.e., a symmetric one where
$\phi=0,N=ev^2/\kappa$ and an asymmetric one where $|\phi|=v, N=0$.  The
particle contents in the symmetry unbroken phase are the complex scalar
$\phi$ and the Dirac fermion $\psi$ with the mass ${e^2v^2/|\kappa|}$
and the neutral scalar $N$, the gauge field $A_\mu$ and the Dirac fermion
$\chi$ with another mass equal to $|\kappa|$. They form four $N=2$
supermultiplets with two degrees of freedom. One supermultiplet consists
of spin 1 and 1/2 and the other three  consist of
spin 1/2 and 0. In the broken phase, we still
have two mass scales
${1\over2}(\kappa^2+4e^2v^2\pm\sqrt{\kappa^2(\kappa^2+8e^2v^2)})$.  The
fields corresponding to these mass eigenstates are obtained as some
combination of the original fields. We have two $N=2$ supermultiplets.
The spin contents in the supermultiplet with the mass $m_+^2 =
{1\over2}(\kappa^2+4e^2v^2 + \sqrt{\kappa^2(\kappa^2+8e^2v^2)})$ are one
with spin 1, two of spin 1/2 and one of spin 0. The spins in the other
supermultiplet with $m_-^2={1\over2}(\kappa^2+4e^2v^2
-\sqrt{\kappa^2(\kappa^2+8e^2v^2)})$ will be that of the $m_+$ with all
the signs changed. The brief analysis of the above supermultiplet
structures is done in the appendix.

Now, let us briefly review the structure of self-dual vortices in the
Maxwell-Chern-Simons theory. The bosonic Lagrangian ${\cal L}_B$ in
Eq.(\ref{Linfty})
is enough for the classical solution. In this theory, there are two degenerate
ground states as mentioned above. It is known that topological
solitons exist in the asymmetric phase with the asymptotic behavior
\begin{equation}
N(\vec r) \rightarrow 0, \quad |\phi(\vec r)|
\rightarrow v \quad \hbox{\rm as} \quad r\rightarrow \infty
\label{asymptop}
\end{equation}
and a quantized flux $\Phi = \pm {2\pi\over e}n$ ($n$: positive integer).
Nontopological solitons also exist in the symmetric phase. We will
consider only the topological vortex for simplicity.
These vortices satisfy the field equations. Especially, they satisfy the
Gauss law constraint
\begin{equation}
\partial_i F^{i0} +\kappa F_{12} + eJ^0 = 0 \label{Gauss}
\end{equation}
with $ J^0 = -i(\phi^*D^0\phi -D^0\phi^*\phi) $.  Integrating over the
whole space then tells us that a configuration with the magnetic flux
$\Phi$ carries the electric charge $Q_E \equiv \int d^2 x
J^0 =-{\kappa\over e} \Phi$.  In this theory it has also been
shown\cite{MCS} that the energy of the configuration is bounded from below
by the relation
\begin{equation}
E\ge ev^2|\Phi| = 2\pi v^2 n , \label{boundinequality}
\end{equation}
and is saturated if the configurations
satisfy the following `self-duality' equations
\begin{eqnarray}
&&(D_1\pm iD_2)\phi \equiv D_\pm \phi =0 , \nonumber\\
&&F_{12} \pm(e|\phi|^2 +\kappa N -ev^2) =0 , \nonumber\\
&&A^0\mp N=0 , \label{self-dual}
\end{eqnarray}
together with the Gauss law (\ref{Gauss}).  The upper (lower) sign
corresponds to a positive (negative) value of the magnetic flux $\Phi$.
Whenever we need the explicit choice of the self-dual background field
configuration, we will choose the upper sign corresponding to the vortex.

In the case of $\kappa=0$, we may consistently set $A^0 = N = 0$ and
Eq.(\ref{self-dual}) will become the self-duality equations for
Landau-Ginzburg vortices\cite{Bogo}.
On the other hand, in the case of $\kappa \rightarrow \infty$, we have
instead
\begin{equation}
A^0 = ({\kappa\over2e^2}) {F_{12}\over |\phi|^2}, \label{a0:infty}
\end{equation}
and the self-duality equations reduce to those of Ref.\cite{CSHiggs},
viz.,
\begin{eqnarray}
&&D_\pm \phi=0 , \nonumber\\ &&F_{12} \pm
{2e^3\over\kappa^2}|\phi|^2(|\phi|^2 -v^2)=0 .
\label{s-d:infty}
\end{eqnarray}

We can take the spherical ansatz for those classical vortices with
vorticity $n$ on top each
other
\begin{equation}
\phi = f(r) e^{in\theta}, \qquad eA^i =\epsilon_{ij}{x_j\over r^2}(a(r)-n).
\end{equation}
The functions are related as
\begin{equation}
a(r) = r{d\over dr} \ln f(r),
\end{equation}
and can be solved using the Bogomolyi equation. Now consider the angular
momentum in the presence of the classical vortices given by the spherical
ansatz.
With the Gauss law now containing the fermion charge density in $J^0$, the
angular momentum from the bosonic fields in
(\ref{ang_bos}) can be written as
\begin{eqnarray}
J_B  &=&  \int d^2 x \epsilon_{ij} x^i \Bigl[\partial^0 N \partial^j N
+D^0\phi^*\partial^j\phi +\partial^j\phi^*D^0\phi
+F^{0k}F^j{}_k \nonumber \\
&&+A^j(-\partial_k E^k +\kappa F_{12} +e\bar\psi\gamma^0\psi) \Bigr] .
\label{ang_bosmod}
\end{eqnarray}
The leading contribution to the angular momentum in (\ref{ang_bosmod}) for
the vortices will be from the
background classical vortex configuration $J_{cl}$ and from the fermion
zero modes $\Delta J_0$.
$J_B \sim J_{cl} +\Delta J_0$.
For vortices with the spherical ansatz,
these are given by \cite{MCS}
\begin{equation}
J_{cl} = -{\pi\kappa\over e^2}n^2
\end{equation}
and
\begin{equation}
\Delta J_0 = -\int d^2x a(r) \psi^\dagger\psi .
\label{anospinfer}
\end{equation}

The fermionic field contribution to the angular momentum in
Eq.(\ref{ang_fer}) under the vortex background becomes
\begin{equation}
J_F = \int d^2x [
\psi^\dagger(-i\partial_\theta +a(r)-{1\over2}\sigma_3-n)\psi
+\chi^\dagger(-i\partial_\theta-{1\over2}\sigma_3)\chi] .
\label{fermodespin}
\end{equation}
The total contribution from the fermion modes will then
become
\begin{equation}
J_F +\Delta J_0 = \int d^2x (\psi^\dagger, \chi^\dagger)
\left[ -i\partial_\theta
-{1\over2} \left( \begin{array}{cc}
\sigma_3 &0\\0 &\sigma_3 \end{array} \right)
-\left( \begin{array}{cc}
n &0\\0 &0 \end{array} \right) \right]
\left( \begin{array}{c} \psi\\ \chi \end{array} \right) . \label{angfersum}
\end{equation}
This will be used in the next section.

\section{Quantization}\label{sec:3}
Now, we want to quantize the theory in the soliton sector. The general
procedure
of the quantization of fields around the soliton has been well
developed \cite{solitonquantize}.  We
decompose the fields around the classical vortex configuration as
\begin{equation}
 \Phi = \Phi_{cl} + \delta\Phi .  \label{decomp}
\end{equation}
The field $ \delta \Phi$ is
the fluctuating modes around the classical vortex configuration
$\Phi_{cl}$.
Plugging eq.(\ref{decomp}) into the Lagrangian, we
have the Lagrangian in the form of
$$ {\cal L} = {\cal L}_{cl} + {\cal L}_{(2)} + {\cal L}_{int} .$$
Here, ${\cal L}_{cl}$ is the same as the bosonic
Lagrangian in Eq.(\ref{Boson}) except that all the fields are the
classical vortex
configuration.  This is the tree level contribution to the Lagrangian
coming from the classical vortex configuration.
${\cal L}_{(2)}= {\cal L}_{B}^{(2)} + {\cal L}_{F}^{(2)}$
is the quadratic piece in terms of the fluctuation fields.
For general value of $\kappa$, the quadratic part of the bosonic
fluctuations is given by
\begin{eqnarray}
{\cal L}_{B}^{(2)}&=&-{1\over 4} \delta F_{\mu\nu} \delta F^{\mu\nu}
 +{1\over4}\kappa\epsilon^{\mu\nu\lambda}\delta F_{\mu\nu} \delta A_\lambda
 -|D_\mu \delta\phi|^2  -e^2 \delta A_\mu^2|\phi|^2  \nonumber \\
& &- ie \delta A_\mu (\delta\phi^* D_\mu \phi +\phi^* D_\mu \delta\phi)
 -{1\over2} (\partial_\mu\delta N)^2  -{1\over2} (\phi^* \delta\phi
       +\phi\delta \phi^* +{\kappa\over e}\delta N)^2\nonumber \\
& & -e^2(|\phi|^2 +{\kappa\over e} N - v^2) |\delta\phi|^2
 -e^2(\delta N)^2 |\phi|^2-e^2 N^2 |\delta\phi|^2
 -e^2 N\delta N(\phi^* \delta\phi +\phi\delta \phi^*),
\label{Bosonquad}
\end{eqnarray}
and that of the fermionic
fluctuations  by
\begin{eqnarray}
{\cal L}_{F}^{(2)} &=&  i\bar\psi \gamma^\mu D_\mu \psi
 + i\bar\chi \gamma^\mu \partial_\mu  \chi  +{\kappa}
   \bar\chi\chi \nonumber\\
&&-i\sqrt{2}e(\bar \psi \chi \phi -\bar\chi \psi \phi^*)
 +eN\bar\psi\psi . \label{FermionN2}
\end{eqnarray}
All the bosonic fields above is the classical background.
Terms in ${\cal L}_{int}$ are
the higher order interaction terms.

The equations of motion for the fluctuating fields around the self-dual
vortices can be obtained by
varying the quadratic piece of the Lagrangian in the above.
First, the equations of motion for the bosonic fluctuations are
\begin{eqnarray}
(\partial_\mu \partial^\mu -2e^2|\phi|^2) \delta A^\nu
-\partial^\nu\partial_\mu\delta A^\mu
-\kappa\epsilon^{\mu\nu\lambda}\partial_\mu\delta A_\lambda
-ie(\phi^* \tensor{D}^\nu \delta \phi +\delta \phi^*
     \tensor{D}^\nu\phi ) &=& 0, \nonumber \\
(-\partial_t^2+ D_-D_+ -2ie A^0\partial_0) \delta\phi
-ie (D_-\phi)\delta A_+ -\kappa e \phi \delta N  +2eA^0\phi(\delta
A^0-\delta N) \qquad && \nonumber \\
-ie\phi  \partial_\mu \delta A^\mu
-e^2\phi(\phi^* \delta \phi +\phi \delta \phi^*)
&=& 0, \label{eqmotgen} \\
(\partial_\mu \partial^\mu  -2e^2|\phi|^2-\kappa^2) \delta N
-e(\kappa + 2e  N) (\phi^* \delta \phi +\phi\delta \phi^*)
&=&0. \nonumber
\end{eqnarray}

Time independent solutions of these equations are the zero modes.
The bosonic zero mode fluctuations may also
be obtained by considering the variation of the self-duality equations
(\ref{self-dual}) around the given classical vortex
configuration as given in Ref.\cite{LeeMinRim}.
Among the zero modes, we eliminate those related with the gauge
transformation by imposing the gauge fixing condition.
The index theorem or its variants
can be used
to count the number of bosonic zero modes satisfying the
equations of motion
and the gauge fixing condition.
The bosonic zero modes for the $\kappa =0$
\cite{vortexindex}, those
for the $\kappa = \infty$ case \cite{JLW90}
and those for a general value of $\kappa$ were studied
\cite{LeeMinRim}.
The results are that in the background of a topological vortex
configuration with vorticity $n$, there exist $2n$ bosonic zero modes for
any value of $\kappa$.
They correspond to the collective coordinates associated with the
vortices.  The quantization of the bosonic zero modes will give rise to
the excitation of the collective coordinates e.g.,the momentum.
This is consistent with the interpretation of these zero modes as being
related to translation of individual vortices.

The general time dependent modes can be quantized as usual to describe the
bosonic particle excitations around the vortices.

We now turn to the fermion fields. The Dirac equation for fermion fields
around the vortex is
\begin{equation}
i \partial_0 \left( \begin{array}{c}
\psi\\ \chi \end{array} \right) = H_F \left( \begin{array}{c}
\psi\\ \chi \end{array} \right), \label{Diraceqgen}
\end{equation}
where the Hamiltonian $H_F$ is given by
\begin{eqnarray}
H_F &=& \left( \begin{array}{cc}
-i\gamma^0\vec\gamma\cdot\vec D + e(-\gamma^0 N+A^0) &
\sqrt{2}ie\gamma^0\phi\\
-\sqrt{2}ie\phi^*\gamma^0 &
-i\gamma^0\vec\gamma\cdot\vec \nabla -\gamma^0\kappa
\end{array} \right) \nonumber \\
&=&
-i\left( \begin{array}{cccc}
0 &D_+&-\sqrt{2}e\phi &0\\ D_- &2ieN&0&\sqrt{2}e\phi\\
\sqrt{2}e\phi^*&0&-i\kappa&\partial_+ \\
0&-\sqrt{2}e\phi^*&\partial_-& i\kappa
\end{array} \right).  \label{Hamiltoniangen}
\end{eqnarray}
The background fields in the above are for vortex configuration
corresponding to the upper sign in Eq.(\ref{self-dual}).

Solutions of the Dirac equation of fermions around the vortex in the above
equation can be decomposed as
\begin{equation}
\Psi =\sum a_i\Psi_{i}^{0}
+\sum_{\omega} b \Psi_{+\omega} +\sum_{\omega} d^\dagger \Psi_{-\omega}
\label{ferdecomp}
\end{equation}
Here $\Psi_{i}^{0}$ are the zero modes and $\sum_{\omega} b \Psi_{+\omega}$
($\sum_{\omega} d^\dagger \Psi_{-\omega}$) are the
positive (negative) energy solutions.
The fermionic zero modes are analyzed in Ref.\cite{Ferzeromode} using the
index theorem and also the relation between the bosonic zero modes and
the fermion zero modes.
There are $2n$ fermion zero modes around the vortex
configuration with winding number $n$ as in the case of bosonic zero modes.
This is deeply related with the $N=2$ supersymmetry
of the theory~\cite{Ferzeromode}.

The
quantization of the zero modes of the fermions will be relevant to the
multiplet contents of the vortices.
To do this, we need to know the angular momentum of the zero modes.
The quantum mechanical angular momentum operator $\cal J$ can be read from
the field theoretical expression in Eq.(\ref{angfersum}) as
\begin{equation}
{\cal J} =
 \left( \begin{array}{cc}
-i\epsilon_{ij} \partial_j -{1\over2}\sigma_3-n & 0 \\
0 & -i\epsilon_{ij} \partial_j -{1\over2}\sigma_3
\end{array} \right) ,
\label{angmomqm}
\end{equation}
It is straigtforward to show that the angular momentum operator $\cal J$
commutes with the Hamiltonian $H_F$.
We decompose the modes into angular momentum eigenstates.
The general mode with the angular
momentum quantum number $j$ is given by
\begin{equation}
\left( \begin{array}{c}
\psi_\uparrow \\ \psi_\downarrow \\ \chi_\uparrow \\ \chi_\downarrow
\end{array} \right)
= \left( \begin{array}{c}
h_1(r)e^{i(j+{1\over2}+n)\theta} \\
h_2(r)e^{i(j-{1\over2}+n)\theta} \\
h_3(r)e^{i(j+{1\over2})\theta} \\
h_4(r)e^{i(j-{1\over2})\theta}  \end{array} \right) e^{-i\omega t},
\end{equation}
The Dirac equation for this mode becomes
\begin{equation}
\omega \left( \begin{array}{c}
h_1(r)\\ h_2(r)\\ h_3(r)\\ h_4(r)\\  \end{array} \right)
= -i \left( \begin{array}{cccc}
0 &\partial_r-{a+j-1/2\over r} & -\sqrt{2}ev f &0 \\
\partial_r+{a+j+1/2\over r} &2ieN&0&\sqrt{2}evf\\
\sqrt{2}evf &0&-i\kappa& \partial_r-{j-1/2\over r} \\
0&-\sqrt{2}ev f &\partial_r+{j+1/2\over r} &i\kappa
\end{array} \right)
\left( \begin{array}{c}
h_1(r)\\ h_2(r)\\ h_3(r)\\ h_4(r)\\  \end{array} \right) \label{eqmodej}
\end{equation}

We first consider the fermion zero modes in the $\kappa\rightarrow 0$
limit. In that case we have two sets of equations.
\begin{equation}
\left( \begin{array}{cc}
\partial_r-{a+j-1/2\over r} & -\sqrt{2}ev f  \\
-\sqrt{2}evf &\partial_r+{j+1/2\over r}
\end{array} \right)
\left( \begin{array}{c}
h_2(r)\\  h_3(r)  \end{array} \right) = 0,    \label{firsthalf}
\end{equation}
and
\begin{equation}
\left( \begin{array}{cc}
\partial_r+{a+j+1/2\over r} & \sqrt{2}ev f  \\
\sqrt{2}evf &\partial_r-{j-1/2\over r}
\end{array} \right)
\left( \begin{array}{c}
h_1(r)\\  h_4(r)  \end{array} \right) =0 . \label{secondhalf}
\end{equation}
The second set of equations (\ref{secondhalf}) does not allow any
normalizable solution.
The first set of equations (\ref{firsthalf}) can be combined as
\begin{equation}
\left(
\partial_r^2 +{\partial_r\over r} -{(j-1/2)^2\over r^2} -2e^2v^2f^2
\right) ({h_2\over \sqrt{2}evf}) = 0 . \label{squaredh2}
\end{equation}
The asymptotic behavior of $h_2(r)$ in
(\ref{squaredh2}) will be proportional to $e^{-\sqrt{2}ev r}$.
Near the origin, $h_2$ behaves as
\begin{equation}
h_2 \sim r^n (A_1 r^{j-{1\over2}} + A_2 r^{-(j-{1\over2})}).
\end{equation}
We expect single solution matching
the boundary condition at infinity with two free parameters at the
origin.
To get the regular solution at the origin with two free parameters, $j$ is
restricted to half integral values in
$-n+{1\over2} \le j \le n-{1\over2}$.
The value $j=n+{1\over2}$ is discarded since corresponding
$\psi_\downarrow$
\begin{equation}
\psi_\downarrow \sim (A_1r^{2n} +A_2 ) e^{2in\theta}
\end{equation}
has bad behavior at $r=0$.
For each of the above $2n$ solutions of $\psi_\downarrow$ ($h_2$), the
function $\psi_\uparrow$ ($h_3(r)$) is
determined through
Eq.(\ref{firsthalf}). Hence we have $2n$ independent zero modes. This
result agrees with that in Ref.\cite{Ferzeromode} based on the index
theorem.

For the general value of $\kappa$, we can get two coupled 2nd order
differential equations of $h_2$ and $h_4$ from (\ref{eqmodej}).
\begin{equation}
\left(\partial_r^2 +{\partial_r\over r}-{(j-1/2)^2\over r^2}
 -2e^2v^2f^2\right) ({h_2\over \sqrt{2}evf}) +i\kappa h_4
=0,             \label{h2h4genkappa1}
\end{equation}
and
\begin{equation}
\left(\partial_r^2 +{\partial_r\over r}-{(j-1/2)^2\over r^2}
-\kappa^2 -2e^2v^2f^2\right)  h_4
-i(\kappa+2eN)2e^2v^2f^2({h_2\over \sqrt{2}evf}) =0. \label{h2h4genkappa2}
\end{equation}
The remaining functions $h_3$ and $h_1$
can be determined by $h_2$ and $h_4$ through the 3rd line and 1st line in
the Dirac equation in (\ref{eqmodej}). The general solutions of
Eqs.(\ref{h2h4genkappa1}) and (\ref{h2h4genkappa2}) are expected to have
four free parameters, to be adjusted to the boundary conditions at the
origin and infinity.
Near $r\sim 0$, the regularity of the solution gives us only three free
parameters for $-n+{1\over2} \le j \le n-{1\over2}$.
The leading orders in the power series expansions in that range of the
angular momentum are given by
\begin{equation}
h_2 \sim r^n (A_1 r^{j-{1\over2}} + A_2 r^{-(j-{1\over2})}),\quad
h_4 \sim B r^{|j-{1\over2}|}.  \label{regorgfin}
\end{equation}
If the angular momentum is out of the above range, then we have at most
two free parameters.
In the asymptotic region, among the four parameters, two will conrrespond
to the unphysical divergent solutions and only two free parameters will
show up in the convergent
solutions  as
\begin{equation}
h_2 \sim C_1 e^{-m_1r} + C_2 e^{-m_2r},\quad
{1\over i\kappa} h_4 \sim {m_1{}^2-2e^2v^2\over \kappa^2} C_1 e^{-m_1r}
 + {m_2{}^2-2e^2v^2\over \kappa^2} C_2 e^{-m_2r} ,\label{integinffin}
\end{equation}
where $m_1{}^2$ and $m_2{}^2$ are eigenvalues of the following mass matrix
\begin{equation}
\left( \begin{array}{cc}
2e^2v^2 & \kappa^2  \\
\kappa^2+2e^2v^2 &2e^2v^2
\end{array} \right).
\end{equation}
Matching the solutions to have the regularity in (\ref{regorgfin}) at the
origin and the integrability in (\ref{integinffin}) will then leave us
only one free parameter
that comes from the homogeneity of the differential equations.
In other words, we have $2n$ solutions, one for each $j$ in the range of
$-n+{1\over2} \le j \le n-{1\over2}$. Note that we do not expect any
solution if $j$ is not in the above range,
since we have less parameters in the
power series solutions near the origin.
Specifically, for $n=1$, we have two modes with
$j=\pm{1\over2}$.

Based on this analysis, we quantize the theory. For simplicity, we
consider the case of single
vortex with the winding number $n=1$.
Multivortex case can be similary done when they are widely separated.  The
quantization $b$ and $d$ in Eq.(\ref{ferdecomp}) (with their conjugates)
for nonzero modes are the same as
that in the vacuum
sector and these modes describe the fermions around the soliton.
We have two fermion zero
modes with the angular momentum
$\pm{1\over2}$.
The quantization for these modes will be
\begin{equation}
 \{a_i, a_j^\dagger\} = \delta_{ij}, \qquad (i,j = 1,2).
\label{quantferzen1}
\end{equation}
The subscript 1 represents for $j=-{1\over2}$ mode and 2 for $j={1\over2}$.
According to Jackiw and Rebbi's interpretation \cite{JackiwRebbi}, the
soliton states will be
degenerate due to the fermion zero modes and
the quantum multiplet structure and the spin contents of the vortices will
form a representation of the algebra relations in equation
(\ref{quantferzen1}).
We will then have four
degenerate soliton states from the fermion zero mode algebra in
(\ref{quantferzen1}).
\begin{equation}
|-->, \quad |+-> = a_1^\dagger |-->, \quad |-+> = a_2^\dagger |-->,\quad
|++> = a_1^\dagger a_2^\dagger |--> . \label{vortexstates:deg}
\end{equation}
The algebra in Eq.(\ref{quantferzen1}) is not the same as the $N=2$ SUSY
algebra with the central charge described in
the appendix. The discrepancy comes from the fact that
only the mode corresponding to $j=-{1\over2}$
can be obtained
by supertranslation of the vortex configuration. To see this, note that we
can always get one fermion zero modes by the supersymmetry tranformation
in Eq.(\ref{susytransf}) to the classical bosonic vortex background. We
then get one fermion zero modes $\Psi_1^{(0)}$ proportional to
\begin{equation}
\psi_\uparrow = 2\sqrt{2}F,\quad \psi_\downarrow =\sqrt{2} iD_-\phi
\quad
\chi_\uparrow = 2iG, \quad
\chi_\downarrow = -2\partial_-N \label{ferzeexp}
\end{equation}
It is straightforward to check that the spin of this zero mode is
$j=-{1\over2}$ using the angular momentum operator in (\ref{angmomqm}).
Hence this is
the zero mode corresponding to $a_1$ in (\ref{quantferzen1}).
We now write the supercharge in (\ref{supercharge}) in two component form
\begin{equation}
Q= \int d^2 x \left( \begin{array}{c}
\sqrt{2}(D_0\phi^*-iF^*)\psi_\uparrow -i(\partial_0 N+f_0
-iG)\chi_\uparrow +\sqrt{2}(D_-\phi)^*\psi_\downarrow
-i(\partial_+ N +f_+)\chi_\downarrow \\
\sqrt{2}(D_0\phi^*+iF^*)\psi_\downarrow -i(\partial_0 N+f_0
+iG)\chi_\downarrow -\sqrt{2}(D_+\phi)^*\psi_\uparrow
+i(\partial_- N +f_-)\chi_\uparrow
 \end{array} \right) \label{superchargeupdown}
\end{equation}
Note that the down component of the supercharge
with the background
self-dual bosonic fields vanishes. The nonvanishing upper component
$Q_\uparrow$ of the supercharge becomes
\begin{equation}
 2 \int d^2 x \left( -\sqrt{2}iF^*\psi_\uparrow -G\chi_\uparrow
+{1\over\sqrt{2}}(D_-\phi)^*\psi_\downarrow -if_+\chi_\downarrow\right) .
\label{superchargeup}
\end{equation}
The algebra of the supercharge will then be that of $N=2$ with the central
charge in the appendix.
The nonvanishing supercharge is in the form of
$Q_\uparrow = \int d^2 x \Psi_1^{(0)} \Psi $. From the
orthogonality of the modes, this is proportional to $a_1$ and so the
operator
$a_2$ corresponding to the other zero mode anticommutes with the
supercharge.
In other words, among the two
independent supertranslations to the self-dual background configurations,
one from $Q_\downarrow$ acts trivially and only the other one from
$Q_\uparrow$ will give
us the fermionic zero mode corresponding to $j=-{1\over2}$.
The other zero mode is not obtained from supersymmetry.  The
algebra between $a_1$ and
$a_1^\dagger$ is the same as the
$N=2$ SUSY algebra with the central term in the appendix and will be
realized as a doublet state.
On the other hand, the doublet representation of the
algebra from $a_2$ and $a_2^\dagger$ will transform as a
singlet under the $N=2$ SUSY.  Hence the above four degenerate solition
states will form two sets of $N=2$ supermultiplets rather than one. One
supermultiplet will be by $|-->$ and $|+->$ with angular momentum
$J_{cl}$ and $J_{cl}-1/2$ respectively. The other one is by $|-+>$ and
$|++>$ with angular momentum
$J_{cl}+1/2$ and $J_{cl}$ respectively.  Here $J_{cl}$ is the leading
contribution to the angular
momentum from the classical bosonic field configuration of
self-dual vortices.

We now calculate the fermion number of the soliton by taking the
expectation value of the fermion number operator for the degenerate
soliton states.
\begin{equation}
<\pm\pm |{1\over2}\int d^2x [\Psi(x)^\dagger, \Psi(x)] |\pm\pm >
=\pm{1\over2}\pm{1\over2} +\eta(H_F) \label{fernumform}
\end{equation}
The constant pieces differing on the vortex structures are from the
fermion zero modes and $\eta(H_F)$, the so-called $\eta$-invariant, given by
\begin{equation}
\eta(H_F) = {1\over 2} (\sum_{\omega>0} -\sum_{\omega<0}) .
\label{etainv:def}
\end{equation}
is from the nonzero modes.

For the Landau-Ginzburg model, i.e., $\kappa=0$ case, there exists a
constant matrix
$
\left( \begin{array}{cc}
\gamma^0 & 0  \\
0 & -\gamma^0 \end{array} \right)
\label{constmtx}
$ that anticommutes with the Hamiltonian. This matrix
matches the positive
energy solution with the negative energy solution with the norm and the
density of states preserved, making the value $\eta$ to be $0$. This can
be also shown easily by direct
evaluation of $\eta$ using the method in Ref.\cite{NiemiSemenoff}, for
example.  Then from Eq.(\ref{fernumform}), the four degenerate states of
the vortex (\ref{vortexstates:deg}) carry fermion number $-1$, 0, 0, 1,
respectively.

The Hamiltonian (\ref{Hamiltoniangen}) for the general value of $\kappa$
no longer has such a
structure. Niemi and Semenoff developed a method to calculate $\eta$ for
such a general case \cite{NiemiSemenoff} which are based on the works of
Atiyah, Patodi and Singer\cite{index}. Following them, let us introduce
one parameter
family of Hamiltonian $H(\tau)$ which interpolates the
Hamiltonian $H_0\equiv H_F(\kappa=0)$ (when $\tau = -\infty$) and
the Hamiltonian $H_{F}$ (when ${\tau = \infty}$).
The Dirac operator $D_\tau\equiv i\gamma^0(\partial_\tau-H_\tau)$
is defined on the extended manifold $\cal M$ of $R^1\times D^2$ where
$R^1=\{\tau\}$ and
$D^2$ is a disk of radius $R$ in the usual $x$-$y$ plane.
The result is \cite{NiemiSemenoff},
in the $R\rightarrow\infty$ limit,
\begin{equation}
-{1\over2} \eta(H_F) = {\rm Index}(D_\tau) -{1\over2} \eta(H_0)
+{1\over2} \eta({\rm Re}({P}) ) \label{indexwithbdry}
\end{equation}
Here, ${\rm Index}(D_\tau)$ is the index of $D_\tau$ in the extended
manifold
$\cal M$ and $P$ is the operator defined by projecting the operator
$D_\tau$ onto the boundary of the disk $D^2$  for each
value of $\tau$. For our case, it becomes
\begin{equation}
{\rm Re} P = -i\gamma^0\partial_\tau +i \left( \begin{array}{cc}
\sigma_1\cos\theta +\sigma_2\sin\theta & 0
\\  0 & \sigma_1\cos\theta +\sigma_2\sin\theta \end{array} \right)
\partial_\theta \label{realMproj}
\end{equation}
which is almost free equation. One can easily see that
$\eta({\rm Re} P)=0$ and $\eta(H_0) =0$. Hence we get
$ \eta(H_F) = -2 {\rm Index}(D_\tau)$.
The index of $D_\tau$ is hard to evaluate. Since the index is
an integer, $\eta$ is an even integer in general.
But we expect its value to be zero, since $\eta$ is shown to be zero in
the above when $\kappa =0$.
In other
words, there is no contribution to the fermion number from the
nonzero modes. Then the fermion number of the vortex is the same as that
in the $\kappa=0$ case.

\section{Quantum correction to the mass of the vortex}\label{sec:4}

The quantization of the nonzero modes in Eq.(\ref{ferdecomp}) will
correspond to the particle creation and annihilation around vortices.
The leading mass correction comes
from the quantum fluctuation of these modes. We will get this correction
by comparing the bosonic and fermionic modes.

The quantization of the nonzero modes in Eq.(\ref{ferdecomp}) will
correspond to the mass correction of the vortex and
particle creation and annihilation around vortices. The mass correction comes
from the sum of those mode contributions \cite{SolitonMass,Rajaraman}. We
will get this correction
by comparing the bosonic and fermionic modes.
First, let us calculate the quantum correction to the mass of
the vortex.
The leading quantum correction to the vortex mass
comes from the quantum
fluctuation modes given by
\begin{equation}
\Delta M = \sum \omega_{B} -\sum \omega_{F}
=\int d\lambda \sqrt{\lambda}
({dn_B(\lambda)\over d \lambda} - {dn_F(\lambda)\over d \lambda}).
\label{masscorr}
\end{equation}
The frequency $\omega_{B}$ ($\omega_{F}$) are eigenvalues and $n_B$($n_F$)
are the number of states upto eigenvalue $\lambda$ of the bosonic
(fermionic) fluctuating modes.

We first evaluate the mass correction
for the usual Maxwell theory ($\kappa=0$).
The equations for the fermionic modes around
the self-dual vortex with the positive frequency $\omega_F$ becomes
\begin{equation}
i\omega_F \left( \begin{array}{c} U \\V \end{array} \right)
= \left(\begin{array}{cc} 0 & D_F\\ -D_F^\dagger &0 \end{array}
\right) \left( \begin{array} {c} U\\V\end{array} \right), \label{eqferkzero}
\end{equation}
where
$$ U=\left( \begin{array} {c}
\psi_\uparrow \\ \chi_\downarrow \end{array} \right), \quad\quad
V=\left( \begin{array} {c}
\psi_\downarrow \\ \chi_\uparrow \end{array} \right),
$$
and the Dirac-like operator $D_F$ is defined as
\begin{equation}
D_F =\left(\begin{array}{cc} D_+ & -\sqrt{2} e\phi \\
-\sqrt{2} e\phi^* & \partial_-  \end{array} \right). \label{DiracOpk0}
\end{equation}
Here the bosonic fields $\phi$ and $A_i$ in the covariant derivatives are
the classical background fields of the self-dual vortex.
Apply the Dirac like operator $D_F$ to the equations of fermions in
Eq.(\ref{eqferkzero})
to get
\begin{equation}
 \omega_F^2 \left( \begin{array}{c} U\\V\end{array} \right)
= \left(\begin{array}{cc} D_F D_F^\dagger & 0
\\ 0 & D_F^\dagger D_F \end{array} \right)
\left( \begin{array}{c} U\\V\end{array} \right). \label{fereqsqr}
\end{equation}
The contribution from the fermionic modes to the soliton mass is given by
\begin{equation}
\sum \omega_{F} =\sum\omega_{U} +\sum\omega_{V}. \label{mcorferz}
\end{equation}
Note that if $(U,V)$ is the mode corresponding to $\omega_F$ then $(U,-V)$
is the mode corresponding to $-\omega_F$.
Let us represent ${dn_+\over d\lambda}$ (${dn_-\over d \lambda}$) as the
density of states of the operator of
$D_F D_F^\dagger$ ($D_F^\dagger D_F$).
Then the density of the states for the fermion modes in
Eq.(\ref{eqferkzero})
is half of those
of the 2nd order equation in Eq.(\ref{fereqsqr})
\begin{equation}
{dn_F(\lambda)\over d \lambda} ={1\over2}
({dn_+(\lambda)\over d \lambda}+{dn_-(\lambda)\over d \lambda}),
\label{denfer}
\end{equation}
since only half of the solution of the above equation correspond to the
positive frequency solution in Eq.(\ref{eqferkzero}).

We now turn to the bosonic fluctuations. The equations of motion in
Eq.(\ref{eqmotgen}) for $\kappa=0$ become
\begin{eqnarray}
(\partial_\mu \partial^\mu -2 e^2|\phi|^2) \delta A_+
-\partial_+\left\{ \partial_\mu \delta A^\mu
+ie(\phi^* \delta \phi -\phi \delta \phi^*)\right\}
+2ie(D_- \phi)^{*} \delta\phi &=& 0, \nonumber \\
(-\partial_t^2+ D_-D_+ -2e^2|\phi|^2) \delta\phi
-ie\phi\left\{ \partial_\mu \delta A^\mu
+ie(\phi^* \delta \phi -\phi \delta \phi^*)\right\}
-ie D_{-}\phi \delta A_{+}&=& 0, \nonumber \\
(\nabla^2  -2e^2|\phi|^2) \delta A^0
+{d\over dt} \left\{ \nabla_i\delta A^i
+ie(\phi^* \delta \phi -\phi \delta \phi^*)\right\}
&=&0, \nonumber \\
(\partial_\mu \partial^\mu  -2e^2|\phi|^2) \delta N &=&0. \label{eqbosonzero}
\end{eqnarray}
The gauge field is written in terms of
$\delta A_{+} = \delta A_1 +i\delta A_2$ for
convenience. The classical background fields $\phi$ and $A_i$
appearing in
the above equations
are the configuration for the self-dual vortices
satisfying the self-dual equations ({self-dual}) with $\kappa=0$.
Among the fluctuations satisfying the above equation (\ref{othereq}),
we have to subtract those fluctuation modes corresponding to the gauge
transformation.
For this purpose, we choose the physical gauge condition
as follows.
\begin{equation}
\nabla_i \delta  A^i
+ie(\phi^* \delta \phi -\phi \delta \phi^*) = 0. \label{gauge:k0}
\end{equation}
The above Eq.(\ref{eqbosonzero}) for $\delta A^0$ with this gauge condition
then requires that
\begin{equation}
\delta A^0 = 0, \label{A0:k0}
\end{equation}
since the operator
$\nabla^2  -2e^2|\phi|^2$ is negative definite.
Other equations for the bosonic fluctuations can then be written as
\begin{eqnarray}
(\partial_\mu \partial^\mu -2 e^2|\phi|^2) \delta A_+
+2ie(D_-\phi)^{*} \delta\phi &=& 0, \label{eqApphi} \\
(-\partial_t^2+ D_-D_+ -2e^2|\phi|^2) \delta\phi
-ieD_{-}\phi \delta A_{+} &=& 0, \label{eqphiAp} \\
(\partial_\mu \partial^\mu  -2e^2|\phi|^2) \delta N &=&0. \label{othereq}
\end{eqnarray}
To find the equations for the fluctuation modes corresponding to the gauge
transformations, we take an infinitesimal gauge
transformation $\delta_G \delta A_+ =\partial_+ \Lambda$ and
$\delta_G \delta \phi = ie \Lambda \phi$ to equations (\ref{othereq}).
We then get the equations for the gauge transformation modes satisfying
\begin{equation}
(\partial_\mu \partial^\mu  -2e^2|\phi|^2) \Lambda =0. \label{flucgaug}
\end{equation}
We have to subtract the degree of freedom satisfying this equation.
Since this equation is the same as that of the real scalar field of
$\delta N$ in
Eq.(\ref{othereq}), the contribution of the fluctuation corresponding to
the gauge
degree of freedom $\Lambda$
satisfying Eq.(\ref{flucgaug}) and that of the neutral scalar $\delta N$
in Eq.(\ref{othereq}) cancels out in the
contribution to the correction of the
mass. Hence the only bosonic contribution is from the complex fields
$\delta A_+$
and $\delta \phi$ and the vortex mass correction from bosons
becomes
\begin{equation}
\sum \omega_{B} = \sum\omega_{\delta \phi} +\sum\omega_{\delta A_+}.
\label{masscorbos}
\end{equation}
To get the density of the states we write the equations for $\delta A_+$
and $\delta \phi$ in Eqs.(\ref{eqbosonzero})
in the matrix form as
follows
\begin{eqnarray}
 -\partial_t^2 \left( \begin{array}{c}
\delta\phi \\ {-i\over\sqrt2} \delta A_+ \end{array} \right)
&=& \left(\begin{array}{cc} D_-D_+ -2e^2|\phi|^2 & \sqrt{2}e\phi
\\ \sqrt{2}e\phi^*  & \partial_i{}^2 -2e^2|\phi|^2 \end{array} \right)
\left( \begin{array}{c}
\delta\phi \\ {-i\over\sqrt2} \delta A_+ \end{array} \right)\\
&=& D_F^\dagger D_F
\left( \begin{array}{c}
\delta\phi \\ {-i\over\sqrt2} \delta A_+ \end{array} \right) ,
\label{eqmatrixform}
\end{eqnarray}
where the Dirac operator $D_F$ is defined in Eq.(\ref{DiracOpk0}).
{}From this we see that the density of states for the bosons are $n_+$.
With the result for fermions in Eq.(\ref{denfer}), this gives that
\begin{equation}
{dn_B(\lambda)\over d \lambda}-{dn_F(\lambda)\over d \lambda}={1\over2}
({dn_+(\lambda)\over d \lambda}-{dn_-(\lambda)\over d \lambda}),
\end{equation}
and hence the mass correction becomes
\begin{eqnarray}
\Delta M &=& \sum \omega_{B} -\sum \omega_{F} \\
&=&{1\over2} \int d\lambda \left(
{dn_+(\lambda)\over d\lambda} -{dn_-(\lambda)\over d\lambda}
\right) \sqrt{\lambda}. \label{indexform}
\end{eqnarray}
We can evaluate the integration  by calculating the density of
states from the phase shift. This method is summarized in the appendix C.
and we get zero for the value of the above integration.

Evaluation of the integrand can also be done with the help of the index
formula \cite{vortexindex}.
\begin{equation}
I(z) = {\rm Tr}\left( {z\over z + D_F^\dagger D_F }
 -{z\over z + D_F D_F^\dagger } \right) = \int_0^\infty d\lambda
{z\over z+\lambda} \left(
{dn_+(\lambda)\over d\lambda} -{dn_-(\lambda)\over d\lambda}\right).
\end{equation}
The index is easily calculated and the result is $2n$. For the index to be
independent of $z$, the integrand in the above should be
\begin{equation}
{dn_+(\lambda)\over d\lambda} -{dn_-(\lambda)\over d\lambda}
= 2n\delta(\lambda).
\end{equation}
By plugging this result into the Eq.(\ref{indexform}) we see that the mass
correction to the vortex in $N=2$ model at one loop
level vanishes.
$$ \Delta M = 0  . $$
This result agrees with Ref.\cite{Schmidt}.

As a check for the independence of the choice of the gauge, let us
describe this in the covariant gauge.
We choose the
background gauge by adding  the following gauge
fixing term to the Lagrangian.
\begin{equation}
{\cal L}_{g.f} = -{1\over2} \left\{ \partial_\mu \delta A^\mu
+ie(\phi^* \delta \phi -\phi \delta \phi^*)\right\}^2.
\end{equation}
The equations for the modes in this gauge are then obtained from the
quadratic pieces of
the Lagrangian by adding .
\begin{equation}
{\cal L}_{B}^{(2)} +{\cal L}_{F}^{(2)}
+{\cal L}_{g.f} +{\cal L}_{gh} .
\end{equation}
The ghost Lagrangian from the gauge fixing is given by
\begin{equation}
 {\cal L}_{gh} = \bar c (\partial_\mu\partial^\mu -2e^2|\phi|^2) c .
\end{equation}
The equations for the fermionic modes are the same as before.
For the equations of the bosonic fluctuations, those for
$\delta A_+$, $\delta \phi$ and $\delta N$ are the same as before in
(\ref{eqApphi}) and (\ref{eqphiAp}). The equations for $\delta A^0$ which
is dynamical in this gauge and
the ghost field $c$  are exactly same as
that of field $\delta N$ in (\ref{othereq}).
Hence the contribution from the bosonic fields and ghosts is  given by
\begin{eqnarray}
\sum \omega_{B} &=& \sum\omega_{\delta \phi} +\sum\omega_{\delta A_+}
 +{1\over2}\sum\omega_{\delta A^0} +{1\over2}\sum\omega_{\delta N}
 -\sum\omega_{c} \\
&=&  \sum\omega_{\delta \phi} +\sum\omega_{\delta A_+}. \label{mcorboskinf}
\end{eqnarray}
We have used the fact that the contribution to the mass from the ghost
fields and that from the
field $\delta N$ and $\delta A^0$ cancel out since the  equation for the
ghost fields are the same
as those for fields $\delta N$ and $\delta A^0$ in Eq.(\ref{othereq}).
We have shown that the only contribution comes from $\delta A_+$ and
$\delta \phi$.
Since equations for these fields are the same as those in the Coulomb
gauge, the remaining arguments are the same as before and so we get the
same result for the mass correction for the vortex.

Now consider another extreme limit of the Lagrangian
with $\kappa\rightarrow\infty$.
The equation of motion of the fermion modes are
\begin{equation}
-\partial_0 \left( \begin{array}{c}
     \psi_\uparrow  \\ \psi_\downarrow \end{array} \right)
=D_{(\infty)} \left( \begin{array}{c}
     \psi_\uparrow  \\ \psi_\downarrow \end{array} \right),
\label{eqferkinfty}
\end{equation}
with
\begin{equation}
D_{(\infty)} = \left(\begin{array}{cc} 2i{e^2\over\kappa}|\phi|^2 & D_+ \\
 D_- & -2i{e^2\over\kappa}(2|\phi|^2-v^2)  \end{array} \right) .
\label{Dirac:infty}
\end{equation}
The equations of bosonic fluctuations can be obtained in a straighforward
way from the Lagrangian (\ref{Linfty}).
\begin{equation}
\kappa \epsilon^{\mu\nu\lambda}\partial_\nu \delta A_\lambda
-2e^2|\phi|^2 \delta A^\mu
-ie(\phi^* \tensor{D}^\mu \delta \phi +\delta\phi^*
\tensor{D}^\mu \phi) = 0,
\label{eqA:infty}
\end{equation}
and
\begin{eqnarray}
D_0^2 \delta\phi +ie(2\delta A^0 D_0 +\partial_0\delta A^0)\phi &=&
D_i^2 \delta \phi
-ie (\delta A^i D_i\phi + D_i \delta A^i \phi)
\nonumber \\
&-&{e^4\over\kappa^2}(9|\phi|^4-8v^2|\phi|^2+v^4) \delta \phi
-{e^4\over\kappa^2}\phi^2\delta \phi^*(6|\phi|^2-4v^2) .
\label{eqPhi:infty}
\end{eqnarray}
We choose the gauge for the spatial gauge fields as
\begin{equation}
\nabla_i\delta A^i
+2i{e^3\over\kappa}(2|\phi|^2-v^2) (\phi^* \delta \phi -\phi \delta\phi^*)
= -{e\over\kappa} (\phi^* \partial^0 \delta \phi -\partial^0\delta\phi^*
\phi). \label{gaugefix:infty}
\end{equation}

The zeroth component of Eq.(\ref{eqA:infty}) and the gauge fixing
condition (\ref{gaugefix:infty}) can be combined into the complex equation
(and its complex conjugate)
\begin{equation}
\partial_- \delta A_+ +4i{e^3\over\kappa^2}(2|\phi|^2-v^2) \phi^* \delta\phi
-2i{e^2\over\kappa} |\phi|^2 {\cal Q}
= 2{e\over\kappa}\phi^* \partial_0 \delta \phi,  \label{eqA0mod:infty}
\end{equation}
where
\begin{equation}
{\cal Q} = \delta A^0 +{e\over\kappa}(\phi^* \delta \phi
 +\delta\phi^* \phi).
\label{A0:infty}
\end{equation}
The spatial component of Eq.(\ref{eqA:infty}) can be written as
\begin{equation}
\partial_0 \delta A_+ + 2i{e^2\over\kappa}|\phi|^2 \delta A_+
-2{e\over\kappa} \phi^*  D_+\delta\phi +\partial _+ {\cal Q}  = 0 .
\label{Ai:infty}
\end{equation}
We will show that the gauge fixing condition and the above equations of
motion gives the $\delta A^0$ so that $\cal Q$ in the above equations
is zero, and $\delta A^0$ is fixed as
\begin{equation}
\delta A^0 = -{e\over\kappa}(\phi^* \delta \phi +\delta\phi^* \phi) .
\label{A0:solved}
\end{equation}
To show
this, we first rewrite the equations for the scalar field
(\ref{eqPhi:infty}) using the gauge fixing condition and the equation
(\ref{eqA0mod:infty}) as
\begin{eqnarray}
\partial_0^2 \delta\phi &=& D_- D_+\delta\phi
-4{e^4\over\kappa^2}(2|\phi|^2-v^2)^2 \delta\phi
-ie\delta A_{+}(D_{-}\phi) \\
&&+ie {1\over\phi^*}(|\phi|^2-v^2) \partial_-\delta A_
+ -ie\phi \partial_0 {\cal Q}.  \label{eqPhimod:infty}
\end{eqnarray}
Comparing this equation with the equation obtained by taking the
time derivative of the equation (\ref{eqA0mod:infty})
we get
\begin{equation}
\nabla^2 {\cal Q} \equiv
\nabla^2(\delta A^0 +{e\over\kappa}(\phi^* \delta \phi +\delta\phi^*
\phi)) = 0 ,
\label{LaplaceQ:0}
\end{equation}
hence ${\cal Q}=0$.
The equations (\ref{eqA0mod:infty}) and
(\ref{Ai:infty}) then become
\begin{equation}
\phi^* \partial_0 \delta \phi = {\kappa\over 2e}\partial_- \delta A_+
+4i{e^2\over\kappa}(2|\phi|^2-v^2) \phi^* \delta\phi ,
\label{eqA0gaugemod:infty}
\end{equation}
and
\begin{equation}
\partial_0 \delta A_+ = -2i{e^2\over\kappa}|\phi|^2 \delta A_+
-2{e\over\kappa} \phi^*  D_+\delta\phi . \label{Aimod:infty}
\end{equation}
And they can replace the set of Eqs. (\ref{eqPhi:infty}),
(\ref{eqA0mod:infty}) and (\ref{Ai:infty}).
The imaginary part of the equation (\ref{eqA0gaugemod:infty}) is nothing
but the gauge fixing condition in Eq.(\ref{gaugefix:infty}). We can also
easily
see that the two equations (\ref{eqA0gaugemod:infty}) and
(\ref{Aimod:infty}) with
$\delta A_0$ given as in
Eq.(\ref{A0:solved}) imply the equation of motion of the scalar
fluctuation in Eq.(\ref{eqPhi:infty}). Hence the physically relevant
bosonic fluctuation modes
are described by the above two equations.

To compare these equations with the fermionic modes we write the equations
(\ref{eqA0gaugemod:infty}) and(\ref {Aimod:infty}).
\begin{equation}
-\partial_0 \left( \begin{array}{c}
 -{\kappa\over2e} {\delta A_+\over\phi^*} \\ \delta \phi \end{array} \right)
=D_{(\infty)} \left( \begin{array}{c}
-{\kappa\over2e} {\delta A_+\over\phi^*} \\ \delta \phi \end{array}
\right) .
\label{eqbosematrix}
\end{equation}
Note that this equation has precisely the same form as that for the
fermionic modes
in equation (\ref{eqferkinfty}).
This means that the density of the modes as well as the spectrum of the
bosons are equal to that of fermions.
Hence the quantum correction for the mass of the vortex vanishes identically.

Finally, let us consider the general case with finite $\kappa$.
The equations for the the bosonic fluctuation fields are given in
Eq.(\ref{eqmotgen}).
We have to fix the gauge to eliminate those fluctuations corresponding to
the gauge transformation.
We choose the following background type gauge condition.
\begin{equation}
\partial_i \delta A^i +\kappa {\cal \delta F}
+ie(\phi^* \delta \phi -\phi \delta \phi^*) = -\partial_0\delta N,
\label{gaugecondgen}
\end{equation}
where ${\cal \delta F}$ satisfies the following equation.
\begin{equation}
\partial_0{\cal \delta F} -\epsilon_{ij}\partial_i\delta A_j
-\kappa\delta N
-e (\phi^* \delta \phi +\phi \delta \phi^*) = 0 .
\label{gauge:genkappa}
\end{equation}
The equations of motion in Eq.(\ref{eqmotgen}) with the help of the
gauge fixing condition
in equation (\ref{gauge:genkappa}) can be written as
\begin{eqnarray}
(\partial_0^2 +D_-D_+ -2e^2|\phi|^2) \delta \phi
-2ieA^0 \partial_0 \delta\phi
-ieD_-\phi \delta A_+ -\kappa e\delta N \phi && \nonumber \\
+2eA^0\phi(\delta A^0 -\delta  N)
+ie\kappa\phi \delta {\cal F} &=&0, \label{eqdelphi} \\
(\partial_i ^2 -2e^2|\phi|^2) \delta A^0
- (\kappa^2 +\partial_0^2) \delta N
-e (\kappa+2eN) (\phi^* \delta \phi +\delta \phi^* \phi )
          &=& 0, \label{eqdelA0}\\
(\partial_\mu \partial^\mu -2e^2|\phi|^2) \delta A_+
+ \kappa\partial_+ \delta {\cal F}
+i\kappa(\partial_+ \delta N+\partial_0 \delta A_+)
+2ie(D_-\phi)^* \delta\phi &=& 0, \label{eqdelAplus} \\
(\partial_\mu \partial^\mu  -\kappa^2-2e^2|\phi|^2) \delta N
-e(\kappa  +2e N)(\phi^* \delta \phi +\phi\delta \phi^*)&=&0
\label{eqdelN}
\end{eqnarray}
We compare Eqs.(\ref{eqdelA0}) and (\ref{eqdelN}) to get
\begin{equation}
(\partial_i^2-2e^2|\phi|^2) (\delta A^0-\delta N) =0.
\end{equation}
Then
\begin{equation}
\delta A^0=\delta N. \label{A0equaltoN0}
\end{equation}
This relation reduces (\ref{eqdelphi}) and (\ref{eqdelAplus}) in the
following form.
\begin{eqnarray}
(-\partial_0^2 +D_-D_+ -2e^2|\phi|^2) \delta \phi -2ieN \partial_0\delta\phi
+ie\kappa\phi (\delta {\cal F} +i\delta N) -ie D_-\phi\delta A_+ &=&0,
\label{redphieq} \\
(-\partial_0^2 +\nabla^2 -2e^2|\phi|^2) A_+
+i\kappa\partial_0 \delta A_+ +\kappa\partial_+ (\delta {\cal F} +i\delta N)
+2ie(D_-\phi)^* \delta\phi &=& 0. \label{reApluseq}
\end{eqnarray}
We have a set of equations (\ref{redphieq}),(\ref{reApluseq}) and
(\ref{eqdelN}) which are supplemented by the
gauge condition (\ref{gaugecondgen}) and (\ref{gauge:genkappa}) or
\begin{equation}
(\partial_0+i\kappa) (\delta {\cal F} +i\delta N) +i \partial_-\delta A_+
-2e\phi^*\delta\phi  = 0 \label{gaugered:finkappa}
\end{equation}

We now turn to the Dirac equation given in Eq.(\ref{Diraceqgen}).
We can remove $\psi_\uparrow$ from this equation by using the first line
of that equation
\begin{equation}
-\partial_0 \psi_\uparrow = D_+ \psi_\downarrow
-\sqrt{2}e\phi\chi_\uparrow  .
\end{equation}
Then we have
\begin{equation}
(-\partial_0^2 +D_-D_+ -2e^2|\phi|^2) \psi_\downarrow
-2ieN \psi_\downarrow - \sqrt{2}e (D_-\phi)\chi_\uparrow
+\sqrt{2}ie\kappa \phi \chi_\downarrow = 0, \label{Dirac:gen1}
\end{equation}
and
\begin{equation}
(-\partial_0^2 +\nabla^2 -2e^2|\phi|^2) \chi_\uparrow
+i\kappa\partial_0 \chi_\uparrow  - \sqrt{2}e (D_-\phi)^*\psi_\downarrow
+i\kappa\partial_+ \chi_\downarrow = 0, \label{Dirac:gen2}
\end{equation}
together with the last line of the Dirac equation.
\begin{equation}
\partial_0 \chi_\downarrow -\sqrt{2}e\phi^* \psi_\downarrow
+ \partial_- \chi_\uparrow +i\kappa \chi_\downarrow = 0 .
\label{Dirac:genlast}
\end{equation}
Comparison of bosonic equations (\ref{redphieq}),(\ref{reApluseq}) and
(\ref{gaugered:finkappa}) with fermionic equations (\ref{Dirac:gen1}),
(\ref{Dirac:gen2}) and (\ref{Dirac:genlast})
gives us the relation
\begin{equation}
\psi_\downarrow = \delta\phi,\quad
\chi_\uparrow = {i\over\sqrt{2}}\delta A_+, \quad
\chi_\downarrow = {1\over\sqrt{2}}(\delta {\cal F} +i\delta N).
\end{equation}
Note that this identification was pointed in Ref.\cite{Ferzeromode} for
the zeromode case.
Here we show that the identification holds also for nonzero mode.
The number of fermion states $n_F$ is simply related with the sum of the
phase shifts of $\psi_\uparrow$, $\psi_\downarrow$, $\chi_\uparrow$, and
$\chi_\downarrow$. In the Appendix C, we will show that the sum of phase
shift of $\psi_\uparrow$ and $\chi_\downarrow$ is equal to the sum of phase
shift of $\psi_\downarrow$ and $\chi_\uparrow$.
Therefore
\begin{equation}
n_F = n_{\psi_\downarrow} +n_{\chi_\uparrow}.
\end{equation}
For the bosonic case, the physical degrees of freedoms are $\delta \phi$,
$\delta N$ and $\delta A_+$ where gauge degree of freedom is subtracted.
However Eq.(\ref{A0equaltoN0}) says that
$\partial_0\delta N = i\omega\delta N$ (assuming
the time dependence as $e^{-i\omega t})$ is nothing but
\begin{equation}
\nabla_i \delta A^i -i\kappa {\cal \delta F}
+ie(\phi^* \delta \phi -\phi \delta \phi^*) = -\omega\delta N,
\end{equation}
which is gauge degree of freedom. So we assert that, as in the case of
$\kappa=0$, we would subtract $\delta N$ instead of subtraction the gauge
degree. Then bosonic degree of freedom is given by $\delta\phi$ and
$\delta A_+$. Then
\begin{equation}
n_B = n_{\delta\phi} +n_{\delta A_+}
\end{equation}
and so we get
\begin{equation}
n_F = n_B .
\end{equation}

Therefore there is no mass correction
\begin{equation}
\Delta M = 0 .
\end{equation}

\section{Summary and Discussion}
We have studied various quantum aspects of the $N=2$ supersymmetric
Maxwell Chern-Simons theory.
First we identified the mass spectrum, the spin contents
and the supermultiplet structures of the particles both in the broken and the
unbroken sectors.

Then, we analyzed the vortex sector, the main subject of this paper.
Starting from the canonical angular momentum we evaluated
the leading quantum correction to the classical value of the angular
momentum of the
vortex coming from the fermion zero modes.
For the supermultiplet structure of vortices,
fermion zero modes
play an important role.
The algebra by the
operators of the fermion zero modes around the winding number $n=1$ vortex
is larger than that of the $N=2$ SUSY algebra with the central charge.
They provide two supermultiplets
with the relative spin difference half, rather than single supermultiplet.
This is in contrast with the case in the monopoles in the 3+1 dimension or
kinks in the 1+1 dimension.
The fermion number of
the vortex is also calculated.

Leading quantum correction to the mass of the vortices is calculated
separately depending whether $\kappa = 0$, $\kappa = \infty$
or finite $\kappa$. The mass correction can be obtained either through
the index theorem or by comparing the modes between bosons and fermions.
In all cases, we do not see any mass correction.

Self-duality is deeply related with the underlying supersymmetry. Here we
have considered only $N=2$ supersymmetric model. For the model with $N=1$
supersymmetry that allows self-dual vortices is not treated. We expect
mass correction in this model since we do not see any simple way of
matching the bosonic and fermionic contributions.

In the models with the self-duality, the mass of the
vortex will be simply related with the magnetic flux of the objects at the
tree level. An interesting question is whether this so-called Bogomolnyi
bound will still be saturated at the quantum level.
There are some models in 2 and 4 dimensions known to satisfy the
Bogomolnyi bound at the quantum level \cite{OliveW,susymonopole}.
To show the saturation at the quantum level, Olive and Witten
\cite{OliveW} used the argument that the size of the
supermultiplets for particles and solitons cannot change abruptly by
perturbation.  This argument seems not to
be directly applied in our case. First, we don't have any self-duality for
particles since we don't have any conserved charge for the particles in
the broken sector. In the vortex sector, for the single
vortex with the winding number 1 to be specific, we have four degenerate
vortex states
from two fermion zero modes. These form two irreducible
supermultiplets of size two with the Bogomolnyi bound saturated. If the
bound is not saturated, we might have only one supermultiplet of size
four. Whether the bound is saturated or not,
the total size of the states remains the same.  On the other hand, in
those models in other dimensions mentioned above, the
supermultiplets of the degenerate solitons form  single irreducible
representation of the superalgebra with the bound saturated.
If the bound does not become saturated by the perturbation, we
need more states, which is unlikely.
This is the difference between our models and
those in other dimensions.
This difference is related to the
fact that all the fermion zero modes in our case do not come from the
supersymmetric transformation of the vortex for winding number one unlike
those other models.

One way to check the saturation of the bound at the quantum
level is to calculate directly the quantum corrections in
Eq.(\ref{boundinequality}).
We have shown that the leading quantum correction to the mass vanish. The
magnetic flux in the right hand side of Eq.(\ref{boundinequality}) may not
get any quantum
correction. We still need the quantum correction to the coupling constants
in the presence of the vortex background to check the validity of the
quantum Bogomolnyi bound.
This is a quite interesting open problem.
For some quantities for the
particles in the vacuum sector, there exist some perturbative calculations
\cite{Effective}.

\acknowledgements
This work was supported in part by the Korea Science and Engineering
Foundation and also by the Basic Science Research Institute Program,
Ministry of Education, 1994, Project No. BSRI-94-2441.  We are grateful to
Professor Choonkyu Lee for his concern and encouragement. B.-H.L would
like to thank Professor G. Semenoff for helpful discussions.

\appendix
\section{Spins}

In this appendix, we will describe how to determine the spin of elementary
excitations. The
spin of the fermion coupled with the CS field was considered in
Ref.\cite{Jackiwspin}. It was found that
the fermions carry spin $\pm 1/2$ and the sign of spin is determined by
the sign of the mass term in the Lagrangian. The spin of vectors in the
case of unbroken Maxwell CS gauge theory was considered in
Ref.\cite{C-Sspin}
while that of broken CS theory considered in Ref.\cite{DeserYang}.
In the following, we will describe how to determine the spin of gauge field
of the Maxwell CS gauge theory
in the broken phase.

We start from the
following Lagrangian
\begin{equation}
{\cal L} = -{1\over 4} F_{\mu\nu} F^{\mu\nu}
+ {\kappa\over 4} \epsilon^{\mu\nu\lambda} F_{\mu\nu} A_\lambda
-{1\over2} \mu^2 A_\mu A^\mu
\label{SpinLag}
\end{equation}
This may be considered  coming from the spontaneously broken
theory of the Maxwell-Chern-Simons Higgs in the unitary gauge. The
canonical variables of this system are
$A_i$ and $\pi_i= F_{0i} +{\kappa\over 2} \epsilon_{ij} A^j$. We separate
the longitudinal and transverse components using the following
identification.
\begin{eqnarray}
A^i &=&\epsilon_{ij} \hat\nabla_j \varphi -\hat\nabla_i \chi \\
\pi_i &=&\epsilon_{ij} \hat\nabla_j \pi_\varphi -\hat\nabla_i \pi_\chi
\label{decompmodes}
\end{eqnarray}
with the abbreviation
$ \hat \nabla_i = {\nabla_i}/\sqrt{-\nabla^2}$.

With these new degrees of freedom, $\varphi$, $\chi$, $\pi_\varphi$, and
$\pi_\chi$, the Hamiltonian density for the system in Eq.(\ref{SpinLag})
is written as
\begin{eqnarray}
{\cal H} &=& {1\over2} ( \pi_\varphi +{\kappa\over2}\chi)^2
+ {1\over2}( \pi_\chi -{\kappa\over2}\varphi)^2
+{1\over2}(\nabla_i\varphi)^2 \nonumber \\
 &&+{1\over2}{1\over\mu^2} \sqrt{-\nabla^2} ( \pi_\chi
+{\kappa\over2}\varphi)^2 + {1\over2}\mu^2(\varphi^2 +\chi^2).
\label{decompHam}
\end{eqnarray}
In the derivation of (\ref{decompHam}), we solved the Gauss law for $A^0$
as
\begin{equation}
A^0 = {1\over\mu^2}(\nabla_i\pi_i +
{\kappa\over2}\epsilon_{ij}\nabla_iA_j) ={1\over\mu^2}\sqrt{-\nabla^2}
(\pi_\chi +{\kappa\over2}\varphi) .
\label{A0solve}
\end{equation}
By varying the Hamiltonian, we can write the equation of motion as
\begin{equation}
 \partial_t^2 \left( \begin{array}{c}
\varphi \\ {\pi_\chi/\kappa}  \end{array} \right)
= \left(\begin{array}{cc} \nabla^2 -\mu^2-{\kappa^2\over2}& \kappa^2\\
 \mu^2+{\kappa^2\over4} &
 \nabla^2 -\mu^2-{\kappa^2\over2} \end{array} \right)
\left( \begin{array}{c}
\varphi \\ {\pi_\chi/\kappa}  \end{array} \right) \label{decompeq}
\end{equation}
with
\begin{equation}
\pi_\varphi +{\kappa\over2}\chi = \dot \varphi , \qquad
(\mu^2+{\kappa^2\over4})\chi +{\kappa\over2}\pi_\varphi = -\dot \pi_\chi
\end{equation}
Note that usual notions of variable and conjugate momentum for $\chi$ and
$\pi_\chi$ has been reversed. The mass matrix in the right hand side of
(\ref{decompeq}) has two eigenvalues.
\begin{equation}
m_\pm^2 = \mu^2+{\kappa^2\over2}
  \pm |\kappa|\sqrt{\mu^2+{\kappa^2\over4}}
\end{equation}
Now, we consider the generaters of the Poincare algebra. The Hamiltonian
is the integration of its density (\ref{decompHam}). The momentum is
\begin{equation}
P^i = \int d^2 x \left[ \varphi \nabla_i \dot\varphi
+{1\over \mu^2} ({\pi_\chi}+{\kappa\over2}\varphi)
\nabla_i ({\dot\pi_\chi+{\kappa\over2}\dot\varphi})\right]
\end{equation}
The angular momentum is
\begin{equation}
J = -\int d^2 \epsilon_{ij} x^i \left[\dot\varphi \nabla_j \varphi
+{1\over \mu^2} ({\dot\pi_\chi+{\kappa\over2}\dot \varphi})
\nabla_j ({\pi_\chi+{\kappa\over2}\varphi}) \right]
\end{equation}
This means that $\varphi$ and $\pi_\chi$ behaves like spin-0 fields.
However the boost generators have infrared problem which was expected from
the decomposition and removing of this infrared divergence will fix the
spin, as first noted in Ref.\cite{C-Sspin}
After some manipulation, we have the following expression for the boost
generator
\begin{equation}
B^i = \int d^2 x\left[ x^i {\cal H}
+\kappa \dot\varphi\epsilon_{ij} {\nabla_j \over -\nabla^2}\varphi
-\dot\varphi\epsilon_{ij}{\nabla_j \over -\nabla^2}
(\pi_\chi+{\kappa\over2}\varphi) -(\dot\pi_\chi+{\kappa\over2}\dot\varphi)
\epsilon_{ij} {\nabla_j \over -\nabla^2}\varphi \right]
\end{equation}
The equation of motion (\ref{decompeq}) can be rewritten as
\begin{equation}
 \partial_t^2 \left( \begin{array}{c}
\varphi \\ {(\pi_\chi +{\kappa\over2}\varphi) / \mu}  \end{array} \right)
= \left(\begin{array}{cc} \nabla^2 -\mu^2-{\kappa^2}& \kappa\mu \\
 \kappa\mu &
 \nabla^2 -\mu^2 \end{array} \right)
 \left( \begin{array}{c}
\varphi \\ {(\pi_\chi +{\kappa\over2}\varphi) / \mu}  \end{array} \right)
\label{decompeq2}
\end{equation}
We can diagonalize the matrix in the right hand side by the new fields
$\xi$ and $\eta$
\begin{equation}
 \left( \begin{array}{c}
 \xi \\ \eta  \end{array} \right)
= \left(\begin{array}{cc}
 \mu\kappa/ N_+ & (m_-^2-\mu^2) /N_+ \\
 (m_+^2-\mu^2) /N_- & \mu\kappa/ N_- \end{array} \right)
 \left( \begin{array}{c}
\varphi \\ {(\pi_\chi +{\kappa\over2}\varphi) / \mu}  \end{array} \right)
\label{redefvar}
\end{equation}
with two suitable normalization constants $N_+$ and $N_-$. With the new
degrees of freedom $\xi$ and $\eta$ the generators of the Poincare algebra
have the following form
\begin{eqnarray}
H &=&{1\over2}\int d^2 x \left[ \dot\xi^2 +\dot\eta^2 +(\nabla_i \xi)^2
+(\nabla_i \eta)^2 +m_+^2\xi^2 +m_-^2\eta^2 \right] \\
P_i &=& \int d^2 x \left[\xi \nabla_i \dot\xi
  +\eta \nabla_i \dot\eta \right] \\
J &=& -\int d^2 x x^i \epsilon_{ij} \left[\dot \xi \nabla_j \xi
  +\dot\eta \nabla_j \eta \right] \label{angmom:decompfields} \\
B^i &=& \int d^2 x \left[ x^i {\cal H}
+m_+\dot\xi\epsilon_{ij}{\nabla_j\over-\nabla^2} \xi
-m_-\dot\eta\epsilon_{ij}{\nabla_j\over-\nabla^2} \eta \right]
\end{eqnarray}
with
$m_\pm =\sqrt{m_\pm^2}=\sqrt{\mu^2+{\kappa^2\over4}}\pm{\kappa\over2} $.
Note that in all formulars, $\xi$ and $\eta$ have the same
contribution except for $B_i$ where the signs of infrared singular terms
are opposite. By direct application of the argument in
Ref. \cite{C-Sspin}, one can show that after removing the
infrared singularity the angular momentum generator
(\ref{angmom:decompfields}) has additional
terms which are propotional to $m_+\over|m_+|$ and $-{m_-\over|m_-|}$
respectively. From those terms, we
can determine spin of $\xi$ and $\eta$ as $+1$ and $-1$.

\section{Supermultiplets}

In this appendix we will describe the supermultiplet structures. We will
consider both the particle
and vortex supermultiplets. The structure will depend on whether we
have the central charge or not. In the case of the vortices or the
symmetry unbroken sector of the vacuum, we will have the central charges
while in the particle spectrum in the symmetry broken sector we have no
central charges. For simplicity we start from
the SUSY alsgebra without the central charge
\begin{equation}
\{Q_A^\alpha, Q_B^\beta\} = 2\bar\sigma_\mu^{\alpha\beta} \delta_{AB} P^\mu
\label{susyalnocent}
\end{equation}
with
\begin{equation}
\bar\sigma_\mu =(I,\sigma_3,\sigma_1)
\end{equation}
We work in the Majorana represention and $Q^\alpha$ corresponds to the real
spinor in $SL(2,R)$ in this appendix.
The index $A,B$
denotes the indices for the extended algebra. In the case of $N=1$, we
have only one supercharge $Q_\alpha$.
In the massive case (this is sufficient in our
analysis), we can take the rest frame and the algebra becomes
\begin{equation}
\{Q^\alpha, Q^\beta\} = 2 \delta^{\alpha\beta} M
\end{equation}
Introducing $ a={1\over 2\sqrt{M}} (Q^1 -iQ^2)$,
$ a^\dagger={1\over 2\sqrt{M}} (Q^1 +iQ^2)$ , we get the algebra
\begin{equation}
\{a,a^\dagger\} = 1 \quad \{a,a\} = 0 \quad \{a^\dagger,a^\dagger\} = 0
\end{equation}
We can construct the Clifford vacuum $|\Omega(j)>$ with spin $j$ by setting
$a|\Omega(j)>=0$.  The angular momentum operator will then be realized as
$J = j a^\dagger a$ in this representation. We will then have two states,
i.e., $|\Omega(j)>$ with spin $j$ and
$a^\dagger|\Omega(j)>$  with spin  $j+{1\over2}$.
To verify this last fact, we note
\begin{equation}
[J, Q^\alpha] = {1\over2} Q_\alpha
\end{equation}
and hence
\begin{equation}
[J, a]= - {1\over2} a, \quad [J, a^\dagger]= {1\over2} a^\dagger.
\end{equation}
We now turn to the case of $N=2$ without the central charge.
The indices $A$ and $B$ will run from 1 to 2.
In this case, we can rewrite the algebra in Eq.(\ref{susyalnocent}) as
\begin{equation}
\{Q^\alpha, Q^{*\beta}\} = 2\bar\sigma_\mu P^\mu
\end{equation}
with the complex notation
$Q^\alpha={1\over\sqrt{2}}(Q_1^\alpha-iQ_2^\alpha)$.
By defining
\begin{eqnarray}
a_1&=&{1\over 2\sqrt{M}} (Q^1 -iQ^2), \quad
a_1^\dagger={1\over 2\sqrt{M}} (Q^{*1} +iQ^{2*}),\quad \nonumber \\
a_2&=&{1\over 2\sqrt{M}} (Q^1 +iQ^2), \quad
a_2^\dagger={1\over 2\sqrt{M}} (Q^{*1} -iQ^{2*})
\end{eqnarray}
we get the following algebra;
\begin{equation}
\{a_1,a^\dagger_1\} = \{a_2,a^\dagger_2\} =1
\end{equation}
\begin{equation}
\{a_1,a_2\} = \{a_1^\dagger,a_2^\dagger\}
= \{a_1,a^\dagger_2\} =\{a_2,a^\dagger_1\} = 0
\end{equation}
and
\begin{equation}
[J, a_1]= - {1\over2} a_1, \quad [J, a_2]= {1\over2} a_2.
\end{equation}
Based on the algebra, we have four states
\begin{equation}
\{ |\Omega(j)>, a_1^\dagger|\Omega(j)>, a_2^\dagger|\Omega(j)>,
a_1^\dagger a_2^\dagger|\Omega(j)> \}
\end{equation}
with angular momentum $(j, j+{1\over2},j-{1\over2}, j)$.

Now consider the supermultiplets in the presence of the central charges.
The central
charge changes the superalgebra
\begin{equation}
\{Q^\alpha, Q^{*\beta}\} = 2\bar\sigma_\mu^{\alpha\beta} P^\mu
-(\sigma_2)^{\alpha\beta} 2C \label{susyalgcentchar}
\end{equation}
with $C=ev\Phi$ in our case. In general $M\ge C$, and
the size of the supermultiplets are the same as the massive
case without the central charge unless the equlity in the above is
saturated, where the size of the
supermultiplets will be reduced.  In the model we are considering, there will
be the central charges for both topological and nontopological vortices
and also for the particles in the symmetry unbroken vacuum sector. In all
these cases
the central charge will be equal to the mass. To see how the
supermultiplets are reduced in these cases, we define
\begin{eqnarray}
a_1&=&{1\over 2\sqrt{M}} (Q^1 -iQ^2), \quad
a_1^\dagger={1\over 2\sqrt{M}} (Q^{*1} +iQ^{2*}), \nonumber \\
a_2&=&{1\over 2\sqrt{M}} (Q^1 +iQ^2), \quad
a_2^\dagger={1\over 2\sqrt{M}} (Q^{*1} -iQ^{2*})
\end{eqnarray}
The superalgebra in Eq.(\ref{susyalgcentchar}) will then become
\begin{equation}
\{a_1,a^\dagger_1\} = 0, \qquad \{a_2,a^\dagger_2\} =1 .
\end{equation}
Other anticommutators vanish. The operators
$a_1 (a_1^\dagger)$ should then be realized to be zero.
Therefore we have only one pair of creation and annihilation operators and
hence two states (not four) $\{ |\Omega(j)>$ and
$a_2^\dagger|\Omega(j)>$ with angular momentum $j$ and $j-{1\over2}$.

\section{Phase shift analysis}
In this appendix we will describe the phase shift analysis used for the mass
correction.
First consider $\kappa =0$ case. Rewrite the Dirac equation in
(\ref{eqmodej}) as
\begin{equation}
L \left( \begin{array}{c}
h_1(r)\\  h_4(r)  \end{array} \right)
\equiv \left( \begin{array}{cc}
\partial_r+{a+j+1/2\over r} & f  \\
f &\partial_r-{j-1/2\over r}
\end{array} \right)
\left( \begin{array}{c}
h_1(r)\\  h_4(r)  \end{array} \right)
= i\omega \left( \begin{array}{c}
h_2(r)\\  h_3(r)  \end{array} \right)
\label{half1}
\end{equation}
and
\begin{equation}
-L^\dagger \left( \begin{array}{c}
h_2(r)\\  h_3(r)  \end{array} \right) \equiv
 \left( \begin{array}{cc}
\partial_r-{a+j-1/2\over r} & - f  \\
-f &\partial_r+{j+1/2\over r}
\end{array} \right)
\left( \begin{array}{c}
h_2(r)\\  h_3(r)  \end{array} \right)
= i\omega \left( \begin{array}{c}
h_1(r)\\  h_4(r)  \end{array} \right)
\label{half2}
\end{equation}
We have set $f'=\sqrt{2} ev f$ and dropped the prime for simplicity.
{}From these first order differential equations, we get the following
2nd order equations.
\begin{equation}
L^\dagger L \left( \begin{array}{c}
h_1(r)\\  h_4(r)  \end{array} \right)
= \omega^2 \left( \begin{array}{c}
h_1(r)\\  h_4(r)  \end{array} \right)
\end{equation}
and
\begin{equation}
LL^\dagger \left( \begin{array}{c}
h_2(r)\\  h_3(r)  \end{array} \right)
= \omega^2\left( \begin{array}{c}
h_2(r)\\  h_3(r)  \end{array} \right)
\end{equation}
where
\begin{equation}
L^\dagger L =   \left( \begin{array}{cc}
-\partial_r^2-{\partial_r+a'\over r} +{(a+j+1/2)^2\over r^2} +f^2 &
 0 \\
0 &
-\partial_r^2-{\partial_r\over r} +{(j-1/2)^2\over r^2} +f^2
\end{array} \right)
\end{equation}
\begin{equation}
LL^\dagger =   \left( \begin{array}{cc}
-\partial_r^2-{\partial_r-a'\over r} +{(a+j-1/2)^2\over r^2} +f^2 &
 2\partial_r f  \\
 2\partial_r f &
-\partial_r^2-{\partial_r\over r} +{(j+1/2)^2\over r^2} +f^2
\end{array} \right)
\end{equation}
Note that $LL^\dagger$ can be obtained by changing the sign of $a$ and $j$
in $L^\dagger L$.
The equations for $h_1$ and $h_4$ are decoupled.
\begin{equation}
 \left(
-\partial_r^2 -{{\partial_r+a'}\over r} +{(a+j+1/2)^2\over r^2} +f^2
\right) h_1
= \omega^2 h_1
\end{equation}
and
\begin{equation}
\left(
-\partial_r^2 -{\partial_r\over r} +{(j-1/2)^2\over r^2} +f^2
\right) h_4 = \omega^2 h_4
\end{equation}
The functions will behave near the origin as
\begin{equation}
h_1 \propto r^{|j+1/2+n|}, \qquad h_4 \propto r^{|j-1/2|}
\label{h:nearorigin}
\end{equation}
In the asymptotic region of $r\rightarrow\infty$, the shape of solutions are
\begin{equation}
h_1(r) \sim \alpha_1 J_{|j+1/2|}(\bar\omega r)
 + \beta_1 N_{|j+1/2|}(\bar\omega r)
\end{equation}
and
\begin{equation}
h_4(r) \sim \alpha_4 J_{|j-1/2|}(\bar\omega r)
 + \beta_4 N_{|j-1/2|}(\bar\omega r)
\end{equation}
with $\bar\omega=\sqrt{\omega^2-1}$.
The coefficients of $\alpha$ and $\beta$ will be determined to match the
behavior near the origin (\ref{h:nearorigin}).

The influence of the background vortex will show up as a phase shift
of the functions by comparing with the case with no vortex background
($a=0$,$f=1$).
The phase shifts are given by
\begin{equation}
\tan \delta_j^{(1)} ={\beta_1\over\alpha_1}
\end{equation}
\begin{equation}
\tan \delta_j^{(4)} ={\beta_4\over\alpha_4}
\end{equation}

The other two functions $h_2$ and $h_3$ are determined by
\begin{equation}
\left( \begin{array}{c}
h_2(r)\\  h_3(r)  \end{array} \right)
= {1\over i\omega} L \left( \begin{array}{c}
  h_1(r)\\  h_4(r)  \end{array} \right)
\label{remainingk0}
\end{equation}
In the limit $r\rightarrow\infty$, we have
\begin{eqnarray}
&&\left( \begin{array}{c}
h_2(r)\\  h_3(r)  \end{array} \right)
\sim
{1\over i\omega}\left( \begin{array}{c}
\partial_r h_1+h_4 \\  \partial_r h_4+h_1   \end{array} \right) \nonumber \\
&\sim&
{1\over i\omega}\left( \begin{array}{c}
-\bar\omega\sin(\bar\omega r -|j+1/2|{\pi\over2}-{\pi\over4}-\delta_j^{(1)})
+\cos(\bar\omega r -|j-1/2|{\pi\over2}-{\pi\over4}-\delta_j^{(4)}) \\
-\bar\omega \sin(\bar\omega r-|j-1/2|{\pi\over2}-{\pi\over4}-\delta_j^{(4)})
+\cos(\bar\omega r -|j+1/2|{\pi\over2}-{\pi\over4}-\delta_j^{(1)})
\end{array} \right)
\label{reducedk0}
\end{eqnarray}
First, consider the case with $j>0$.
The phase shift is given by
\begin{equation}
\left( \begin{array}{c}
h_2(r)\\  h_3(r)  \end{array} \right)
\sim
{1\over i\omega}\left( \begin{array}{c}
-\bar\omega\sin(R-\delta_j^{(1)})
+\cos(R-\delta_j^{(4)}+{\pi\over2}) \\
-\bar\omega \sin(R'-\delta_j^{(4)})
+\cos(R'-\delta_j^{(1)} -{\pi\over2})
\end{array} \right)
\label{posjk0}
\end{equation}
with
\begin{equation}
R=\bar\omega r -(j+1/2){\pi\over2}-{\pi\over4},\quad
R'=\bar\omega r -(j-1/2){\pi\over2}-{\pi\over4} =R+{\pi\over2}
\end{equation}
Then the phase shifts become
\begin{equation}
\tan (\delta_j^{(2)} - \delta_j^{(1)}) =
-{\sin(\delta_j^{(4)}-\delta_j^{(1)})\over
 \bar\omega+\cos(\delta_j^{(4)}- \delta_j^{(1)})}
\end{equation}
and
\begin{equation}
\tan (\delta_j^{(3)} - \delta_j^{(4)}) =
-{\sin(\delta_j^{(4)}-\delta_j^{(1)})\over
 \bar\omega-\cos(\delta_j^{(4)}- \delta_j^{(1)})}
\end{equation}
For the case with negative $j$,
the phase shift will be
\begin{equation}
\left( \begin{array}{c}
h_2(r)\\  h_3(r)  \end{array} \right)
\sim
{1\over i\omega}\left( \begin{array}{c}
-\bar\omega\sin(R''-\delta_j^{(1)} )
+\cos(R''-\delta_j^{(4)}+{\pi\over2}) \\
-\bar\omega \sin(R'''-\delta_j^{(4)})
+\cos(R'''-\delta_j^{(1)} +{\pi\over2})
\end{array} \right)
\label{negjk0}
\end{equation}
with
\begin{equation}
R''=\bar\omega r +(j+1/2){\pi\over2}-{\pi\over4},\quad
R'''=\bar\omega r +(j-1/2){\pi\over2}-{\pi\over4} =R''+{\pi\over2}
\end{equation}
The phase shift will then be
\begin{equation}
\tan (\delta_j^{(2)} - \delta_j^{(1)}) =
{\sin(\delta_j^{(4)}-\delta_j^{(1)})\over
 \bar\omega+\cos(\delta_j^{(4)}- \delta_j^{(1)})}
\end{equation}
and
\begin{equation}
\tan (\delta_j^{(3)} - \delta_j^{(4)}) =
{\sin(\delta_j^{(4)}-\delta_j^{(1)})\over
 \bar\omega+\cos(\delta_j^{(4)}- \delta_j^{(1)})} \label{posnegrelk0}
\end{equation}
{}From the above equations, not depending on the sign of $j$
\begin{equation}
\tan (\delta_j^{(2)} - \delta_j^{(1)}) =
-\tan (\delta_{-j}^{(3)} - \delta_{-j}^{(4)})
\end{equation}
hence
\begin{equation}
\delta_j^{(2)} - \delta_j^{(1)} =
-(\delta_{-j}^{(3)} - \delta_{-j}^{(4)})
\end{equation}

So, for fixed $\omega$
\begin{equation}
\sum_j(\delta_{j}^{(2)} + \delta_j^{(3)})
- \sum_j(\delta_{j}^{(1)} +\delta_j^{(4)})
= \sum_j(\delta_{j}^{(2)} - \delta_j^{(1)})
+\sum_j(\delta_{j}^{(3)} -\delta_j^{(4)})
=0
\end{equation}
{}From the relation between the phase shift and the density of states
\cite{Rajaraman}
we get
\begin{equation}
n_+=n_-. \label{nplus:nminus}
\end{equation}

We turn to the case with $\kappa \ne 0$.
The equations for $h_2$ and $h_3$ are
\begin{eqnarray}
 \Bigl(
\partial_r^2 & & +{{\partial_r +a'}\over r} -{(a+j-1/2)^2\over r^2}
+\omega^2-f^2 -2\omega N -{\kappa\over{\omega-\kappa}}f \Bigr) h_2
\nonumber \\
&+&
 \left({1\over {\omega-\kappa}} (\partial_r
+{1\over r} (a+j+{1\over 2}) ) - {2af\over r} \right)  h_3 = 0
\end{eqnarray}
and
\begin{eqnarray}
 \Bigl(
\partial_r^2 &&+{{\partial_r+a'}\over r} -{(j+1/2)^2\over r^2}
+(\omega-\kappa)^2 -(1-{\kappa\over\omega})f^2 \Bigr) h_3
\nonumber \\ && +\Bigl( {{-2af}\over r}
 -{\kappa\over\omega}(\partial_r-{1\over r}(a +j-1/2) )\Bigr) h_2 = 0
\end{eqnarray}
By solving these two coupled equations,  we can get $h_2$ and $h_3$.
$h_1$ and $h_4$ are fixed by
\begin{equation}
h_1 = {1\over i\omega} \left(
(\partial_r -{1\over r} (a+j-{1\over 2}))  h_2 -f h_3 \right)
\end{equation}

\begin{equation}
h_4 = {1\over {i(\omega-\kappa)}} \left( (\partial_r
+{1\over r} (j+{1\over 2})) h_3 -f h_2 \right)
\end{equation}
unless $\omega=0$ and $\omega=\kappa$.

Following the steps for the case of $\kappa=0$, we get the result
\begin{equation}
\delta \psi_\uparrow +\delta \chi_\downarrow =
\delta \psi_\downarrow +\delta \chi_\uparrow ,
\end{equation}
which is similar to Eq.(\ref{nplus:nminus})

There may exists a solution to the Dirac equation in (\ref{eqmodej})
for threshold value of $\omega=\kappa$. But
this gives us a discrete modes which does not contribute to the mass
correction.

\def\ap#1#2#3{Ann.\ Phys.\ (N.Y.) {\bf {#1}}, #3 (19{#2})}
\def\imp#1#2#3{Int.\ Mod.\ Phy.\ {\bf {A#1}}, #3 (19{#2})}
\def\pl#1#2#3{Phys.\ Lett.\ {\bf {#1}B}, #3 (19{#2})}
\def\np#1#2#3{Nucl.\ Phys.\ {\bf B{#1}}, #3 (19{#2})}
\def\prl#1#2#3{Phys.\ Rev.\ Lett.\ {\bf #1}, #3 (19{#2})}
\def\pr#1#2#3{Phys.\ Rev.\ {\bf D{#1}}, #3 (19{#2})}
\def\prp#1#2#3{Phys.\ Report\ {\bf {#1}C}, #3 (19{#2})}
\def\nuovo#1#2#3{Nuovo Cimento {\bf {#1}B}, #3 (19{#2})}
\def\commath#1#2#3{Comm.\ Math.\ Phys.\ {\bf {#1}}, #3 (19{#2})}
\def\ibid#1#2#3{{\bf {#1}}, #3 (19{#2})}

\end{document}